\documentclass[prl, twocolumn]{revtex4}
\usepackage{amsmath}
\usepackage{graphicx}
\usepackage{xcolor}
\def\lsim{\mathrel{\raise3pt\hbox to 8pt{\raise -6pt\hbox{$\sim$}\hss{$<$}}}}

\begin{document}
\bibliographystyle{apsrev}

\title{Possible Origins and Implications of the Shoulder in Reactor Neutrino Spectra }

\author{A. C. Hayes$^1$, J. L. Friar$^1$, G. T. Garvey$^1$, Duligur Ibeling$^{1,2}$, Gerard Jungman$^1$, T. Kawano$^1$,  Robert W. Mills$^3$}
\affiliation{$^1$Los Alamos National Laboratory, Los Alamos, NM, USA  87545}
\affiliation{$^2$Harvard College, 	Cambridge, MA, USA 02138}
\affiliation{$^3$National Nuclear Laboratory, Sellafield, CA20 1PG UK}

\begin{abstract}
We analyze within a nuclear database framework the shoulder observed in the antineutrino spectra in current reactor experiments. We find that the ENDF/B-VII.1 database predicts that the antineutrino shoulder arises from an analogous shoulder in the aggregate fission beta spectra. 
In contrast, the JEFF-3.1.1 database does not predict a shoulder for two out of three of the modern reactor neutrino experiments,
and the shoulder that is predicted by JEFF-3.1.1 arises from $^{238}$U.  
We consider several possible origins of the shoulder, and find possible explanations.
For example, there could be a problem with the measured aggregate beta spectra, or 
 the harder neutron spectrum at a light-water power reactor could affect the distribution of beta-decaying isotopes.
In addition to the fissile actinides, we find that $^{238}$U could also play a significant role in distorting the total antineutrino spectrum.
Distinguishing these and quantifying whether there is an anomaly associated with measured reactor neutrino signals will require new short-baseline experiments, both at thermal reactors and at reactors with a sizable epithermal neutron component.
\end{abstract}
\maketitle
Modern reactor neutrino experiments measuring $\theta_{13}$, 
such as Daya Bay \cite{daya-bay}, RENO \cite{reno}, and Double Chooz \cite{double-chooz}, 
involve detectors both near and far from the reactors. 
The shape and magnitude of the antineutrino spectra emitted from the reactors 
have been measured to high accuracy in the near detectors of both Daya Bay and RENO. 
The Daya Bay near-detector has also provided an absolute determination of the reactor antineutrino flux, 
and this is consistent in magnitude with the previous world average short-baseline reactor neutrino experiments. 
As such, the measured magnitude is consistent with a deficit with respect to the most recent estimates \cite{huber, mueller} of the expected reactor antineutrino flux. 
The absolute magnitude of the RENO flux has yet to be published. 
However, in the near detector of both RENO and Daya Bay the shapes of the measured spectra are not consistent with the antineutrino spectrum predictions \cite{huber, mueller} that we refer to as the Huber-Mueller model. 
Most notably, the measured antineutrino spectra exhibit a significant shoulder relative to the model predictions at antineutrino energies $\sim 5-7$ MeV. 
The spectra measured at Daya Bay, RENO, and Double Chooz all exhibit this shoulder. 
Thus, there are two puzzles associated with measured reactor antineutrino spectra:
(1) the yield in all short-baseline experiments is lower than current models, and 
(2) the shape of the measured spectra deviate from these model predictions. 
However, these two issues are not necessarily related.

In the Daya Bay, RENO and Double Chooz experiments the antineutrinos are measured by 
detecting the positrons produced in inverse beta decay on the protons
 ($\overline{\nu}_e+p\rightarrow n+e^+$) in the detector, 
and the positron energy is reconstructed from the scintillation light created 
by the kinetic energy of the positron and its annihilation. 
The antineutrino spectrum $S(E_\nu)$ emitted from a reactor is determined by \cite{cao} the reactor thermal power ($W_{th}$), 
the energy released in fission by each actinide ($e_i$), 
the fractional contribution ($f_i/F$, $F=\Sigma_if_i$)  
of each actinide to the fissions taking place, and the antineutrino spectrum for each actinide $S_i(E_i)$:
\begin{equation}
S(E_\nu)=\frac{W_{th}}{\sum_i (f_i/F)e_i}\sum_i (f_i/F)S_i(E_\nu) .
\end{equation}
Corrections to Eq.~(1) arise from nuclei with long half-lives that do not reach equilibrium in the reactor, except for very long
burn times.
In addition, there are low-energy antineutrino contributions from the spent fuel. 
Both of these effects must be taken into
account in analyses of reactor neutrino experiments, as discussed for example in refs.~\cite{two1, two2}.
The thermal power and the fission fractions are both functions of time and are supplied by the  
reactor operator, 
while the energy contributing to the thermal power per fission of each actinide ($e_i$) 
is normally taken from refs.~\cite{James, kopeikin,correct}.
 Over most of the observed spectrum at the three experiments the measured shape differs from the predictions, 
and the measurements in the 4-6 MeV prompt-energy ($E_\nu\approx E_{\rm prompt}+0.782$ MeV) shoulder region represent an $\sim 4\sigma$
 deviation from expectation. 
The spectral shape of this shoulder cannot be produced by any standard L/E dependence required of neutrino oscillations, 
sterile or otherwise. Thus, there is a need to investigate the origin of the shoulder within a more detailed nuclear physics framework

In this report we examine the uncertainties in the antineutrino fluxes and consider five possible origins of the shoulder. 
These include:
(1) antineutrinos produced by neutron reactions with reactor (non-fuel) materials; 
(2) consequences of the forbidden nature of the beta decays dominating the antineutrino flux in the shoulder region; 
(3) contributions from $^{238}$U;
(4) potential effects due to a harder neutron spectrum in a pressurized water reactor (PWR);
(5) a problem with the original aggregate beta spectra on which the expected antineutrino flux is based. 
Of these, only (1) can be eliminated as a possible source.
The other four sources could contribute to the shoulder in varying degrees.

Dwyer and Langford \cite{dwyer} examined the database predictions using a  subset of the ENDF/B-VII.1 \cite{endf-7} library. 
They found that the shoulder appears to result from the contribution of a few energetic decays that should have also produced a shoulder in the aggregate beta spectrum. 
Since the release of the ENDF/B-VII.1 library, it has been shown by Fallot {\it et al.}~\cite{Fallot} that updates to the decay library 
that include beta-decay branches from an analysis of total absorption gamma-ray spectroscopic (TAGS) measurements 
\cite{Greenwood} and the  beta spectra of ref.~\cite{Tengblad} are crucial for producing fission-aggregate beta and antineutrino spectra.  The TAGS method is sensitive to low-energy beta decays than are not seen in some other techniques. Missed low-energy decays will lead to excessive strength being assigned to high-energy decays.
 The updates of refs.~\cite{Greenwood, Tengblad} were not included in the 
analysis of Dwyer and Langford \cite{dwyer}. 

Sonzogni {\it et al.} \cite{bnl} included these updated beta-decay data in their analysis using  the JEFF-3.1.1 \cite{JEFF} fission-yield
 and ENDF/B-VII.1 beta-decay libraries. 
In the current work we use the same updated ENDF/B-VII.1 beta-decay library as Sonzogni {\it et al.} \cite{bnl, bnl2},  and we 
 compare the results of using either the JEFF-3.1.1  or the ENDF/B-VII.1 fission-yield libraries. 
As described in ref.~\cite{bnl}, TAGS data \cite{Greenwood} (where available) were used for all nuclei listed in ref.~\cite{bnl2}
for the updated  library, while for all other nuclei
 listed in ref.~\cite{bnl2} the data of ref.~\cite{Tengblad} were used.
In addition for $^{92}$Rb, which is a dominant contributor to the high-energy component of the spectrum, we followed the recommendation of Sonzogni {\it et al.} and used the 
beta-decay spectrum of ref.~\cite{Tengblad}, which corresponds to a $0^-\rightarrow 0^+$ branching ratio of 95\%.
A
 comparison between the old and updated decay libraries for the shapes and magnitudes 
of the aggregate fission beta-decay spectra for the actinides of interest is provided by Fallot {\it et al.}~\cite{Fallot},
where the spectral changes are shown to be significant at all energies, including in the energy region of the shoulder. 
The relative importance of the dominant nuclei contributing to the shoulder for the old and updated libraries is
provided by Dwyer and Langford~\cite{dwyer} and Sonzogni {\it et al.}~\cite{bnl}, respectively.
Sonzogni {\it et al.}~\cite{bnl} found good agreement between their database analysis 
and the Schreckenbach measurements \cite{schreck,schreck2,schreck-238} of the beta-decay aggregate fission spectra, although they did not make direct comparisons between the two 
with the accuracy needed to reveal either the shoulder or the anomaly. 
As in ref.~\cite{bnl}, the current work includes the database evaluations for all fission fragments, 
using modeled spectra \cite{shannon, kawano} for fragments with unmeasured decay spectra.


Following detailed arguments presented below, we conclude that PWR antineutrino fluxes are not known to the accuracy suggested by the current models \cite{huber, mueller}. 
There are two methods for deducing the antineutrino flux. Both start with establishing the underlying beta spectra. 
The first method measures an aggregate beta spectrum and fits it to a number of end-point energies 
to generate an antineutrino spectrum. 
The second tries to assemble the underlying beta and antineutrino 
spectra from the fission yields and decay data for all the fission fragments in a database.  
The current uncertainty \cite{hayes} in converting the measured aggregate beta spectrum to an antineutrino spectrum is about 4\%,
 while the analysis below and and that of Sonzogni {\it et al.}~\cite{bnl} 
leads to the conclusion that the uncertainty in using a database summation is appreciably larger.
The aggregate fission beta spectrum $N_\beta(E_e)$ under equilibrium reactor burning conditions 
for a given actinide is determined by a summation of the beta spectra $S(E_e,Z_i,A_i)$ of the individual beta-unstable 
fission fragments $F_i$ weighted by their cumulative fission yields $Y_{F_i}$:
\begin{equation}
N_\beta(E_e) = \sum_{F_i} Y_{F_i}\, S(E_e,Z_i,A_i) \, .
\label{agg}
\end{equation}
The beta spectrum $S(E)$ for each fragment $(Z_i,A_i)$ summed over all decay branches is normalized to unity: $\int S(E,Z,A)\,dE=1$.
 In all calculations presented here the corrections to beta decay suggested in ref.~\cite{mueller}, 
but using the forms derived in ref.~\cite{hayes}, are included in calculating $S(E_e,Z_i,A_i)$. 
Both the ENDFB/V-II.1 and JEFF-3.1.1 libraries provide cumulative yields $Y_{F_i}$ for all
fission fragments of interest. 
The updated ENDF/B-VII.1 beta-decay library \cite{bnl, bnl2} provides spectra for approximately 95\% of the nuclei appearing in Eq.~(2). 
The remaining 5\% of the fission fragments are modeled \cite{shannon, kawano} by extension 
of the Finite-Range Droplet Model plus Quasi-particle Random Phase Approximation (QRPA). 
The model has been tuned to account for the so-called pandemonium effect \cite{Hardy}  (viz., a very large number of low-energy beta decays to high-lying excited states of the daughter) as well as forbidden transitions, 
and is supplemented by the nuclear structure library ENSDF \cite{ENSDF} where appropriate. 
The model provides a good description of fission-decay heat \cite{decayheat1, decayheat2, decayheat3}, 
and of  TAGS \cite{Greenwood} measurements for individual nuclei. 
In Fig.~(1) we show the relative importance of the modeled spectra for the Daya Bay combination of fissioning actinides.
\begin{figure}
\includegraphics[width=3.0 in]{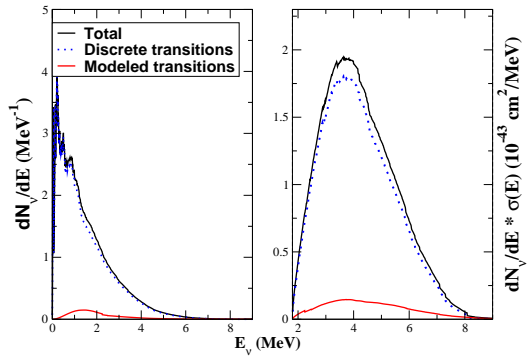}
\caption{Approximately 5\% of the fission fragments have unknown beta-decay spectra that can only be modeled.
The figure shows the measured versus modeled contributions to the total aggregate antineutrino spectrum
for Daya Bay, as predicted using the JEFF-3.1.1 fission yields (left panel). The same spectra folded over the 
neutrino detection cross section are shown in the right panel.}
\end{figure}
In Figs.~(2) and (3) below we take the uncertainty in the total modeled portion of the spectra to be 50\%, which is then added in quadrature with
the uncertainty in the aggregate spectra from the other 95\% of nuclei, the latter being taken from ENDF/B-VII.1.

Figure~2 shows the database predictions for the shape of the antineutrino spectra for Daya Bay \cite{Zhong} and RENO \cite{Kim} relative to 
the Huber-Mueller model \cite{huber, mueller}, and for Double Chooz \cite{buck3} relative to the Huber-Haag model\cite{huber, schreck-238, buck2}. 
The Daya Bay, RENO, and Double Chooz experiments differ in the linear combination of actinides determining the total fissions. 
For Daya Bay the $^{235}$U: $^{238}$U: $^{239}$Pu: $^{241}$Pu fission split is 0.586: 0.076: 0.288: 0.05. 
RENO has not published their fission split, but we took 0.62: 0.12: 0.21: 0.05 from ref.~\cite{Kim}, and the Double Chooz
split to be \cite{buck2} 0.496: 0.087: 0.351: 0.066. 
As can be seen in Fig.~(2), the ENDF/B-VII.1 fission-fragment yields lead to the prediction 
of a shoulder relative to the Huber-Mueller model, but the JEFF-3.1.1 yields do not. 
This striking difference arises because the cumulative fission yields for some nuclei that dominate 
in the shoulder region are different in the two evaluations. 
For several nuclei that contribute to the shoulder-energy region, the TAGS data tend to correct for the pandemonium effect and  suppress the highest-energy beta-decay branches.  
That is, these data include beta-decay branches to previously omitted low-energy transitions, 
which thereby suppress the branching ratios for the highest-energy decays.
This has the effect of reducing the magnitude of the predicted shoulder.
In the case of Double Chooz,
 the JEFF-3.1.1 fission yields do predict a shoulder and this shoulder arises almost entirely from $^{238}$U.
The JEFF-3.1.1 prediction of a shoulder for Double Chooz and not for Daya Bay and RENO
occurs because the former experiment uses the Haag \cite{schreck-238} antineutrino spectrum for $^{238}$U
as opposed to the Mueller spectrum. 
Both ENDF/B-VII.1 and JEFF-3.1.1 predict a shoulder with respect to $^{238}$U alone, regardless of  whether the Haag or the Mueller antineutrino spectrum is used, with the former spectrum producing a shoulder almost twice as big as the latter.

In Table 1 we list the main differences between the fission yields in the two database evaluations.
Rather than discuss the reasons behind each of these differences, we examine the differences for $^{96}$Y,
which is a dominant nucleus contributing to the shoulder region \cite{dwyer, bnl}. 
\begin{table}
\caption{ Dominant nuclei contributing to the shoulder for which the databases disagree on the fission  yields by more than 20\%.
The other dominant nuclei $^{88,91}$Br, $^{92,93,94,96}$Rb, $^{138,140}$I, and $^{144}$Cs have similar yields in both databases.
The relative importance of the contribution of these nuclei in the shoulder region is displayed in refs.~\cite{dwyer, bnl}. }
\vspace{4pt}
\begin{tabular}{l|cc|cc}
Nucleus&\multicolumn{2}{|c|}{JEFF $Y_{F_i}$ (\%)}&\multicolumn{2}{|c}{ENDF $Y_{F_i}$ (\%)}\\\hline
&$^{235}$U&$^{239}$Pu&$^{235}$U&$^{239}$Pu\\\hline
$^{89}$Br&1.36&0.50&1.08&0.35\\
$^{90}$Br&0.49&0.10&0.56&0.25\\
$^{95}$Rb&0.66&0.26&0.77&0.44\\
$^{96}$Y&4.72&2.88&6.0&4.35\\
$^{97}$Y&2.08&1.22&4.89&3.75\\
$^{98}$Y&1.07&0.68&1.92&1.52\\
$^{98m}$Y&1.97&1.87&1.11&1.19\\
$^{100}$Y&0.30&0.21&0.61&0.35\\
$^{134m}$Sb&0.52&0.19&0.36&0.20\\\hline
\end{tabular}
\end{table}
This nucleus has both a 0$^-$ ground state (g.s.) and an 8$^+$ isomeric level; 
 the ground state contributes significantly to the shoulder, while the isomer does not. 
The discrepancy between the evaluated fission yields results from assumptions made about the fission split to the isomer versus the ground state. 
In the ENDF/B-VII.1 (JEFF-3.1.1) evaluation, 90\% (64\%) of the independent fission yield goes to the isomer and 10\% (36\%)
to the g.s. In addition, ENDF/B-VII.1 assumes that the $^{96}$Y isomer gamma decays 100\%  to the $^{96}$Y g.s. \cite{Y96}.
There are no measurements of the isomer and g.s. fission splits. 
A range on the yield of $^{96}$Y can be obtained by assuming 0\% and 100\% splitting to the $^{96}$Y g.s., 
which gives a $^{96}$Y yield range of 3.75-6.05\% for $^{235}$U and 1.86-4.35\% for $^{239}$Pu. 
This range exemplifies the degree of uncertainty in the evaluated fission yields resulting from the modeled feeding to different isomeric states and their subsequent decays.
In obtaining these numbers we note that the g.s. gets a significant contribution to its cumulative yield from the beta decay of $^{96}$Sr.

Within the ENDF/B-VII.1 analysis, the shoulder in the antineutrino spectrum results from a corresponding shoulder 
in the aggregate beta spectrum, and involves the decay of several nuclei, as listed in ref.~\cite{dwyer}. 
The very large over-prediction of the beta and anti-neutrino spectra at energies above about 7.5 MeV in \cite{dwyer} 
is the result of not using the TAGS data.
 With the updated beta-decay library this problem is greatly reduced, but not removed, as can be seen in Fig.~(3).
\begin{figure}
\includegraphics[width=2.8 in]{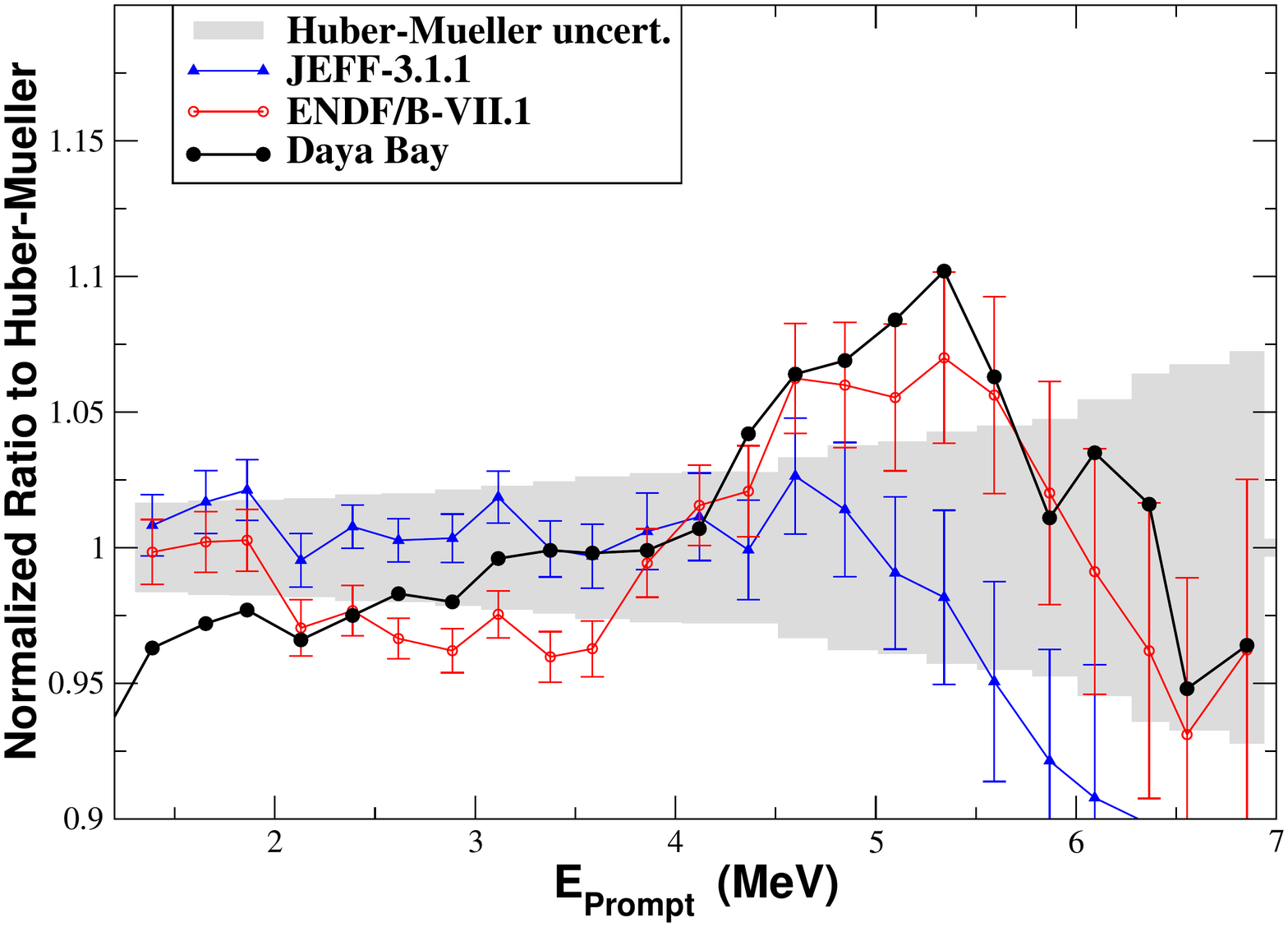}
\includegraphics[width=2.8 in]{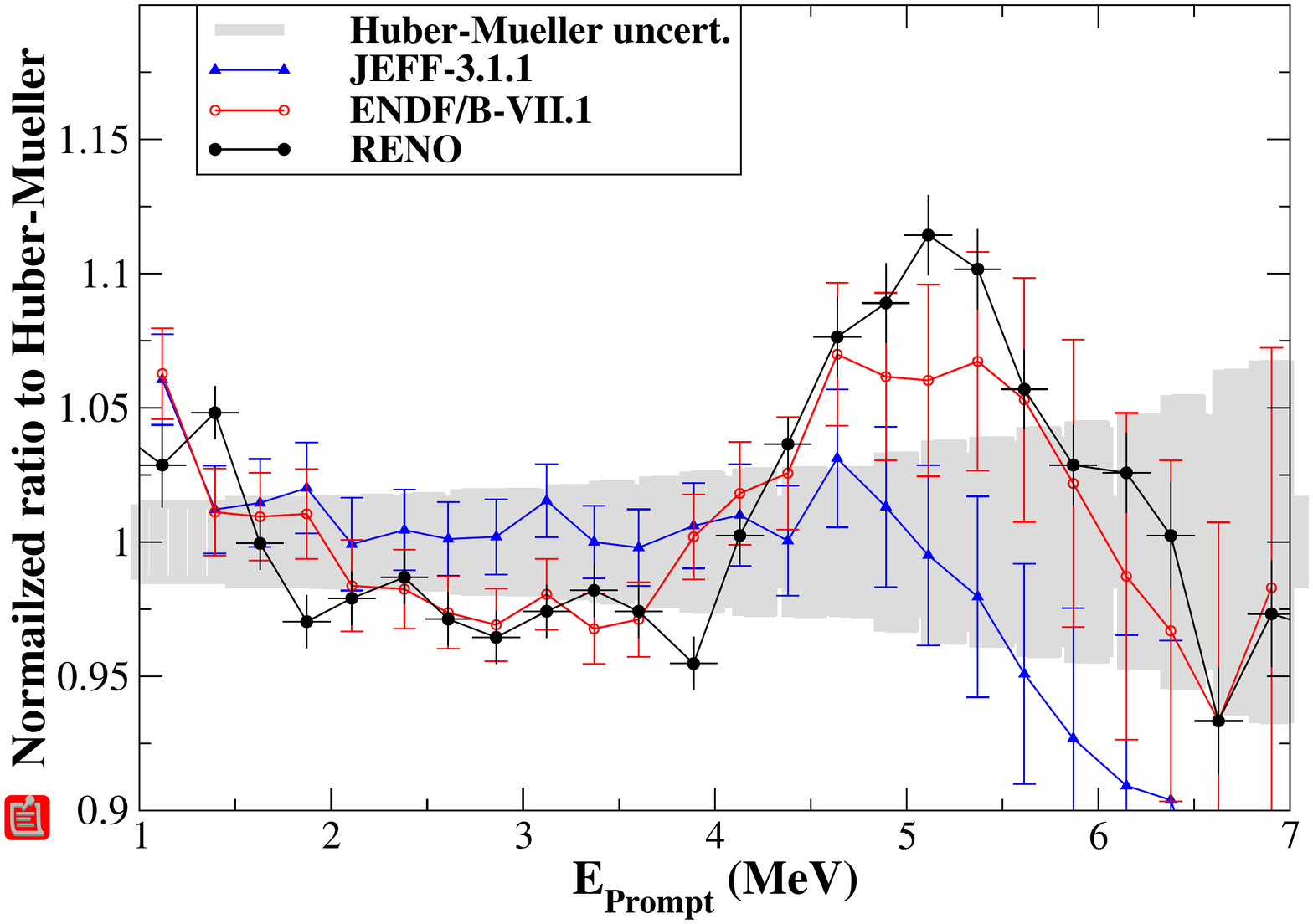}
\includegraphics[width=2.8 in]{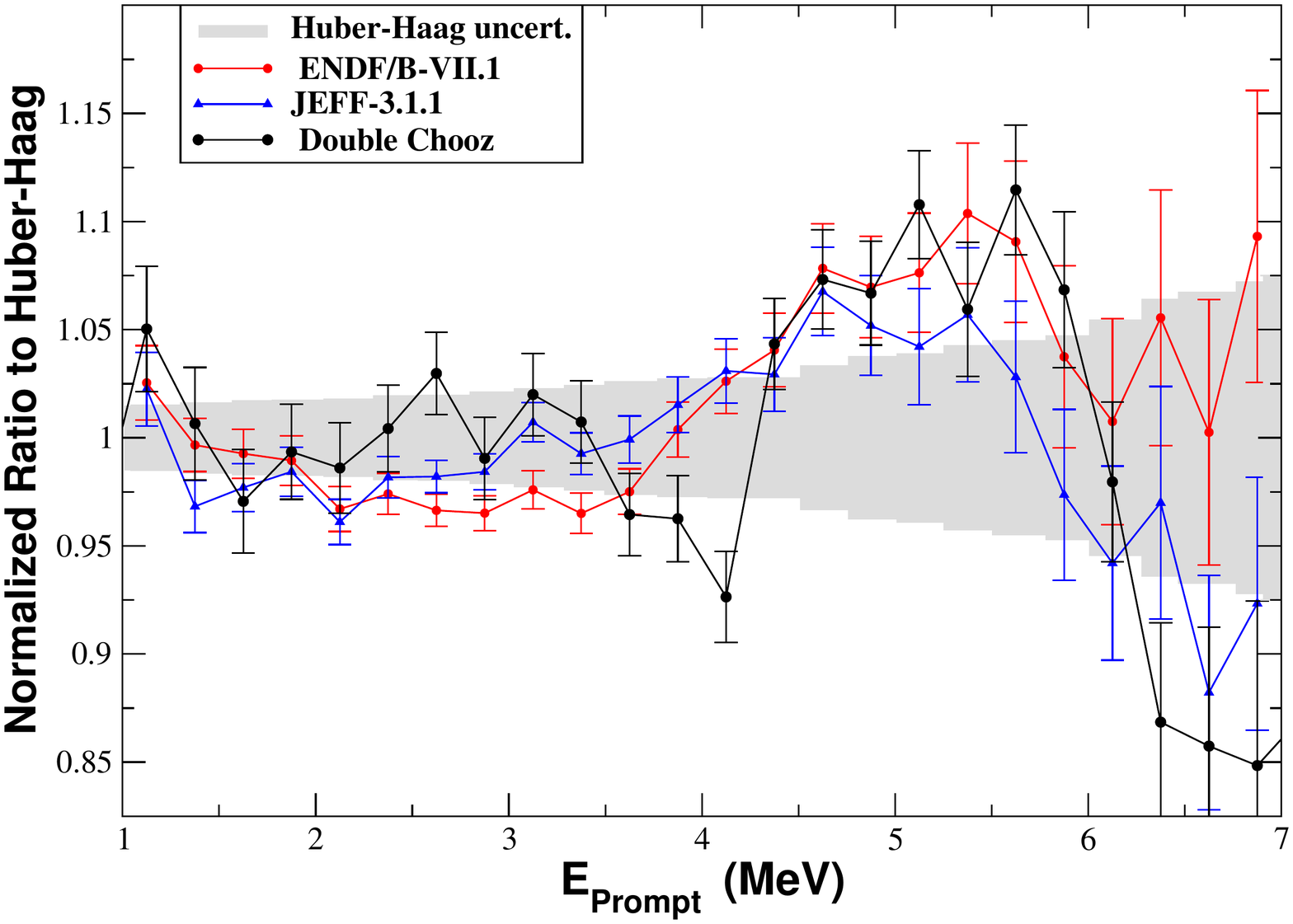}
\caption{The ENDF/B-VII.1 and JEFF-3.1.1 predictions, including the beta-decay database update \cite{bnl2}, for the ratio of the Daya Bay \cite{Zhong}, RENO \cite{Kim} and Double Chooz \cite{buck3} antineutrino spectra to 
the Huber-Mueller or Huber-Haag models \cite{huber, mueller, schreck-238, buck2}, as labeled on by the y-axis. 
In all cases, the spectra are normalized to the same number of detectable antineutrinos in the energy window 
$E_\nu = 2-8$ MeV ($E_\nu\approx E_{\rm prompt}+0.782$ MeV) as the Huber-Mueller (or Huber-Haag) spectra when folded over the antineutrino detection cross section \cite{nu-cross}. 
The database uncertainties shown are only for the beta-decay branches. 
The uncertainties arising from the fission-fragment yields are large, as is evident from the difference between the ENDF/B-VII.1 and JEFF-3.1.1 predictions. 
 The large difference between the two database predictions for the shoulder, particularly for Daya Bay and RENO,
 arises entirely from a difference in the evaluated fission-fragment yields. 
The predicted shoulder for JEFF 3.1.1 relative to the Huber-Haag prediction for Double Chooz arises because the 
Haag prediction for $^{238}$U is appreciably smaller in the shoulder region than JEFF 3.1.1.
}
\end{figure}

We next discuss in detail possible origins of the shoulder. 
In particular, we seek to identify sources that can generate a shoulder in the energy interval 
$4.0 < E_{\rm prompt} < 6.5$ MeV 
that can account for $\sim 2\%$ of the total $E_{\rm prompt}$ spectrum as reported by RENO \cite{Kim}.\\
\noindent {\it 1. Non-fission sources of antineutrinos:} We examined the contribution to the antineutrino
spectrum from neutron-induced reactions in reactor materials other than the fuel.
We used MCNP simulations that are available for all neutron-induced reactions on the coolant, cladding, and structural materials in the NRU CANDU reactor at Chalk River.
We then calculated the expected beta-decay spectrum from the unstable nuclei produced by these reactions.
We found that all of the antineutrinos from this source are well below the energy of the shoulder.
This is consistent with the analysis of ref.~\cite{two2}.
While materials in other reactors may differ in detail from those at the NRU reactor,
none is known to produce a significant number of antineutrinos above 2 MeV, and
we conclude that non-actinide sources of antineutrinos cannot explain the shoulder.

\noindent {\it   2. The forbidden nature of transitions:}
 Several of the beta-decay transitions involving $^{96,98}$Y, $^{90,92}$Rb, and $^{142}$Cs that dominate in the shoulder region  
have a total angular momentum and parity change  that generates no weak-magnetism correction \cite{hayes}. 
This fact was not taken into account in the analyses of Huber, Mueller, or Fallot. 
Above half of the end-point energy in an allowed decay, the weak-magnetism contribution reduces the antineutrino component. 
This is opposite in sign to the other leading corrections \cite{huber, mueller, hayes} that suggested the existence of the reactor anomaly \cite{mention}. 
Thus, the lack of a weak-magnetism correction for $0^+\rightarrow 0^-$ transitions increases the magnitude of the antineutrino flux relative to the Huber-Mueller model. 
A second issue is that the shape factor, $C(E)$, associated with $0^+\rightarrow 0^-$ forbidden transitions \cite{hayes} 
is quite different from the approximation used by  Mueller {\it et al.} \cite{mueller}, who took
 the shape factor for all forbidden transitions to be that for a unique forbidden transition.
A third issue is the lack of a proper finite-size Coulomb correction to the Fermi function
for these transitions \cite{hayes}, where all analyses to-date (including the present one) are  forced to use an approximation.

We calculated the antineutrino spectra with and without taking
the $\Delta J^{\Delta\pi}=0^-$  nature of transitions into account. 
There are two possible shape factors for such transitions \cite{hayes} that affect the spectrum differently,
 which introduces an uncertainty in the shape of the aggregate antineutrino
spectrum.  
Using the shape factor that gives the bigger increase in the antineutrino spectrum and
setting the weak-magnetism term to zero, we found an increase in the shoulder region of less than 1\%. 
We conclude that a proper treatment of forbidden transitions cannot account for a significant fraction
 of the shoulder.

\noindent {\it 3. $^{238}$U as a source of the shoulder:} 
RENO reports that $^{238}$U is responsible for about 12\% of its fissions, while Daya Bay reports only 7.6\%. 
Referring to Fig.~(1), relative to their respective experimentally established base lines
(rather than with respect to Huber-Muller), the RENO shoulder is more than 50\% larger than that observed at Daya Bay.
This raises the question whether $^{238}$U, which was not measured in the original ILL experiments \cite{schreck, schreck2},
 could be causing the shoulder.
Because 
$^{238}$U fissions into isotopes further off the line of stability than $^{235}$U, its antineutrino spectrum is both larger and harder in energy, and   
in the region $E_{\rm prompt}=4-6 $ MeV the $^{238}$U spectrum is almost twice as large as that of $^{235}$U. 
Thus, $^{238}$U contributes about 24\% (15\%) to the total spectrum in the shoulder region for RENO (Daya Bay).
We compared the ENDF/B-VII-1 and JEFF-3.1.1 predictions for $^{238}$U  to Mueller's prediction \cite{mueller}, and
found that {\underline both} databases predict a significant shoulder for $^{238}$U. 
The magnitude of the JEFF-3.1.1 (ENDF/B-VII.1)  shoulder and the percentage contribution to the total antineutrino spectrum suggests that
$^{238}$U could account for 25\% (50\%) of the observed shoulder in RENO and Daya Bay.
To account for the entire shoulder in these two experiments the fast fission-fragment  yields for $^{238}$U 
 dominating the shoulder region would have to be on average about 
a factor of four (two) larger than the JEFF-3.1.1 (ENDF/B-VII.1) evaluations. 
In Double Chooz $^{238}$U accounts for 8.7\% of the fissions, and the Haag spectrum (as opposed to the Mueller spectrum) was taken \cite{schreck-238, buck2}
 as the expected for $^{238}$U. 
JEFF-3.1.1 predicts a  shoulder for Double Chooz and it is  almost entirely due to $^{238}$U.
Thus, we conclude that $^{238}$U could be responsible for a significant fraction of the observed shoulder. 
But without experiments designed to isolate the contributions from each actinide to the shoulder,
$^{238}$U cannot be assumed to be responsible for the entire shoulder.

\noindent {\it   4. The relatively harder PWR Neutron Spectrum:}
 The neutron flux spectra at the PWR reactors used by Daya Bay, RENO and Double Chooz are
 harder in energy than the thermal spectrum of the ILL reactor, and involve considerably larger epithermal components. 
This raises the question whether epithermal neutron contributions to the 
fission of $^{235}$U, $^{239}$Pu and $^{241}$Pu could result in a shoulder in the antineutrino spectrum. 
Studies \cite{epithermal} of energy-dependent variations in the fission-product
yields found clear evidence for significant yield changes 
for nuclei
 in the valley of the double-humped mass-yield curve. For example, the epithermal
 yield (relative to thermal) for the relatively unimportant isotope
 $^{115}$Cd  varies by a factor of 0.5-3.0, depending on the particular energies of epithermal fission resonances.
The effects are much more pronounced in $^{239}$Pu than in $^{235}$U. Resonance-to-resonance
fluctuations cause the {\it average} effect to be small ($\sim$ 4\%) in the energy range $19 < E_n < 61 $ eV for $^{235}$U,
while in $^{239}$Pu the prominent and isolated resonance at 0.3 eV produces a change in the $^{115}$Cd yield of more than a factor of two.
For high-yield fission products, such as $^{96}$Y and $^{92}$Rb, yield changes are not expected to be as large
as for nuclei like $^{115}$Cd, both because of theoretical arguments \cite{wheeler} and because
 the sum of the independent yields is fixed. 
But changes of the order of 20\% are not ruled out.
There have been some experiments to examine \cite{popa, willis, tong}  changes in the fission yields of isotopes that sit at the peaks of the mass distribution, but the results are discrepant: some experiments observe changes \cite{popa, willis} and others do not \cite{tong}.
One issue in considering the hardness of the spectrum is that since epithermal fission-yield effects are observed \cite{epithermal} to be larger
in $^{239}$Pu than in $^{235}$U, the larger shoulder but smaller $^{239}$Pu contribution at 
RENO would seem at odds with the shoulder being solely induced by a harder neutron flux.
However, with no fission-yield measurements (thermal or epithermal) 
for
nuclei that dominate the shoulder region we conclude that 
the hardness of the neutron flux spectra cannot be ruled out as a contributing factor. 
For example, the 0.3 eV $^{239}$Pu fission resonance
plays a much more significant role in PWR reactors than at the ILL reactor.
Thus, a  comparison of  the antineutrino spectrum measured at a very thermal reactor 
with that at a reactor with a sizable epithermal neutron component would be valuable in addressing this issue.
Experimental determinations of the fission yields of the nuclei dominating the shoulder and their variations with 
neutron energy would also be valuable. 

\noindent {\it 5. A possible error in the ILL  beta-decay measurements:}
 As pointed out by Dwyer and Langford \cite{dwyer} the ENDF/B-VII.1 prediction of a shoulder in the antineutrino spectrum in Fig.~(1)
 corresponds to an analogous shoulder in the aggregate beta spectrum. 
In Fig.~(2) we show the absolute ratio of the ENDF/B-VII.1 prediction for the aggregate beta spectrum for $^{235}$U to that of
Schreckenbach \cite{schreck, schreck2}. 
We conclude that the shoulder could be the result of a problem in the measurement or analysis of the beta spectra produced at ILL. 
\begin{figure}
\includegraphics[width=2.8 in]{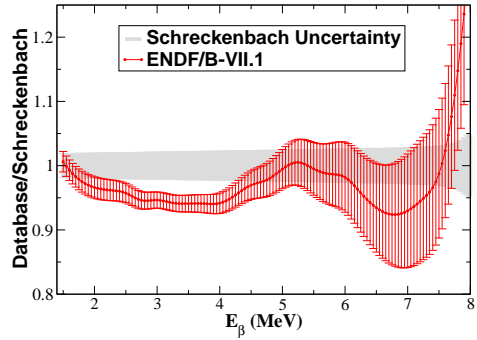}
\caption{.
The absolute ratio of the ENDF/B-VII.1 aggregate beta spectrum for $^{235}$U to that of Schreckenbach \cite{schreck}. 
The shoulder in the energy window $E_\beta \approx 4-6$ MeV corresponds to the same shoulder in the ENDF/B-VII.1 antineutrino spectrum shown in Fig.~(2).}
\end{figure}

Finally, we comment on whether database analyses of the antineutrino spectra provide any insight into the reactor neutrino anomaly \cite{mention}. 
The most important comment is that the 
database uncertainties are too large to draw any conclusions.
Nonetheless, it is noteworthy that in comparing the two 
fission-yield evaluations, the prediction of a shoulder (no shoulder) 
appears to be correlated with the predictions of no anomaly (an anomaly).
Daya Bay observes a shoulder and its measured absolute rate is in excellent agreement \cite{Zhong} with the previous world average.
The ENDF/B-VII.1 prediction for both the shoulder and 
the absolute magnitude of the antineutrino spectrum are close to Daya Bay; that is, relative to ENDF/B-VII.1, Daya Bay sees no anomaly.
In contrast, the JEFF-3.1.1 predictions are closer to the Huber-Mueller model, which would suggest an anomaly.

Both the anomaly and the shoulder could be due to 
(a) a difference in the hardness of the reactor neutron spectrum, 
or (b) a problem with the original aggregate beta-spectra measurement \cite{schreck, schreck2} at the ILL. 
It is also possible that the shoulder and the anomaly are not correlated. 
Answering  these questions is not possible within current theoretical frameworks or from existing data. 
Consequently a new set of reactor experiments is needed at short baselines. 
To address the important issue of the anomaly and the possible existence of a $1$ eV sterile neutrino, detection at 
different distances viewing the same reactor are needed. 
To quantify the role of the neutron spectrum on the shape and magnitude of the antineutrino spectrum, 
one measurement should be carried out at a very thermal reactor and the 
other at a reactor with a considerably harder neutron spectrum. 
The use of highly enriched $^{235}$U fuel has the advantage of 
restricting the resulting antineutrino flux to fragments produced by a single actinide.
In addition to addressing the possible origin of the anomaly and shoulder, 
a detailed measurement of the shape of the $^{235}$U antineutrino spectrum would be very valuable in
shedding light on the differences between the ENDF/B-VII.1 and JEFF-3.1.1 fission yields,  as well as examining the reliability of the ILL measurement of the $^{235}$U spectrum.
On the other hand, if $^{238}$U and/or $^{239}$Pu play a significant role in the anomaly or the shoulder, 
measurements from fuel that is of low enrichment will be needed to reduce these sources of uncertainty.


\begin{thebibliography}{99}
\bibitem{Y96} We note that there is an inconsistency between the ENDF/B-VII.1 fission-yield and electromagnetic-decay data for this
nucleus. The decay library shows that the $^{96}$Y isomer beta decays 100\% to $^{96}$Zr.
The fission-yield library assumes that this isomer electromagnetically decays 100\% to $^{96gs}$Y.
\end{thebibliography}
\end{document}